 \documentclass[doublecol]{epl2} %for 2 columns style without line numbers
\def\be{\begin{equation}}
\def\ee{\end{equation}}

\def\bc{\begin{center}}
\def\ec{\end{center}}
\def\bea{\begin{eqnarray}}
\def\eea{\end{eqnarray}}
\newcommand{\avg}[1]{\langle{#1}\rangle}
\newcommand{\Avg}[1]{\left\langle{#1}\right\rangle}

\title{Statistical mechanics of bipartite $z$-matchings}
\shorttitle{Statistical mechanics of bipartite $z$-matchings} %Insert here a short version of the title if it exceeds 70 characters

\author{Eleonora Krea\v{c}i\'{c}  \inst{1} \and Ginestra Bianconi\inst{2,3} }
\shortauthor{E. Krea\v{c}i\'{c} \etal}

\institute{ 
 \inst{1}Department of Statistics, University of Oxford,  Oxford OX1 3LB, UK\\
  \inst{2}School of Mathematical Sciences, Queen Mary University of London, London E1 4NS, UK\\
  \inst{3}Alan Turing Institute, The British Library, London, UK}
  

\abstract{
The  matching problem has a large variety of applications including  the allocation of competitive resources and  network controllability. The statistical mechanics approach based on the cavity method has shown to be exact in characterising this combinatorial problem on locally tree-like networks. Here we use the cavity method to solve the many-to-one bipartite $z$-matching problem that can be considered to be a model for  the characterisation of the capacity of user-server networks such as wireless communication networks. Finally we study the phase diagram of the model defined in network ensembles.}

\begin{document}

\maketitle

\section{Introduction}

In network science \cite{Laszlo,Newman,Latora,bianconi_2018multilayer} there is increasing interest in combinatorial optimisation problems and message-passing algorithms applied to processes as different as control \cite{Control,Menichetti_PRL,Menichetti_multilayer}, percolation \cite{Lenka_Newman,bianconi2018rare}, percolation on multilayer networks \cite{bianconi_2018multilayer,Phys_Rep,Doro1,Doro2,Radicchi_percolation,Osat} or epidemic spreading \cite{Zecchina1,Zecchina2,Gleeson,Saad}. 
On one hand, this surge of interest is motivated by the efficiency of using statistical mechanics approach \cite{Mezard_Montanari,Weigt} in solving  combinatorial  optimisation problems. On the other hand, it comes as a consequence of the vast realm of applications of these problems and their generalisations. In this paper we characterise the statistical mechanics of a generalised matching problem called $z$-matching that can be interpreted as a model in a system with limited resources.

On an undirected network the matching problem consists of finding the maximum subset of links of the network (the matched links) such that each node is adjacent to at most one link from that subset. This problem has attracted large interest from combinatorics, probability and the computer science communities \cite{KarpSipserER,Aronson,Gale}. For this problem the statistical mechanics approach \cite{Lelarge,Spin_glass,Mezard_Parisi} is very useful and in particular the Belief Propagation algorithm \cite{Lenka_Mezard,Chayes,HJ,HJ2} provides the exact solution as long as the network is locally tree-like.

The matching problem and its generalisations have a variety of applications ranging from wireless networks to network controllability. The generalisation of the matching problem on directed networks has recently been shown \cite{Control} to characterise the network controllability as it identifies the set of driver nodes of the network. Since the matching problem on directed locally tree-like networks is exactly solvable using statistical mechanics methods, this result has opened new perspectives in network controllability. In particular, it has allowed to relate the directed network dynamical properties (controllability) to its structural ones (the properties of its maximum matching). Interestingly in this context it has been shown \cite{Menichetti_PRL} that the key structural property characterising the matching (and hence the network controllability) is the fraction of nodes with low in and out degrees.These results have  also recently been  extended to multilayer network controllability by considering  the relevant extension of the maximum matching problem to a multilayer network maximum matching problem \cite{Menichetti_multilayer}.

More traditionally the matching problem has been defined in spatial networks where nodes have specific positions, or more generally in networks in which each pair of nodes is associated with a distance, the so called marriage problem \cite{Zhang,Zhang_marriage,Caldarelli}. In this setting the generalised matching problem aims at finding the maximum matching that minimises the overall distances between the nodes. This specific model defined over a bipartite network  has a variety of applications \cite{Zhang} in finite resource allocation problems where some providers of service want to optimise the user satisfaction and their own profit.

In this paper, we are focusing on another important variation of the matching problem called the many-to-one $z$-matching on bipartite networks. In this case we consider a bipartite network in which one set of nodes can be matched at most to one link and the other set of nodes can be matched to at most $z>1$ links. 
The many-to-one $z$-matching problem on bipartite networks has  recently become particularly  relevant for characterising  wireless communication networks \cite{Wireless,Alex} in which one has two different sets of nodes -- users and towers. Each tower provides the wireless connection, but can only serve up to $z$ users. The \textit{maximum capacity} of the network is given by the largest number of users that can be served at a time. Here, we determine the Belief Propagation equations to characterise the maximum capacity of any locally tree-like bipartite network and we evaluate the capacity of network ensembles with given degree distribution. Our analysis is based on the use of the Belief Propagation algorithm in the zero-temperature limit. In this way we extend the existing statistical mechanics treatment of the one-to-one maximum matching problem of simple, directed and multilayer networks to the many-to-one matching problem.

\section{The  maximum $z$-matching problem}\label{Sect Max Matching Problem}

Let us consider a bipartite network formed by $N$ users $i=1,\cdots,N$ and $M$ towers $\alpha=1,\cdots,M$ where we assume for simplicity that each tower is connected to at least $z$ users.  A \textit{$z$-matching} is the subset of the set of edges such that each user is adjacent to at most one and each tower is adjacent to at most $z$ edges from the subset. In other words, each user communicates with at most one neighbouring tower and each tower serves at most $z$ neighbouring users. The \textit{size} of a $z$-matching is given by the number of its edges and the {\it maximum capacity} of the network is given by the size of the largest possible $z$-matching. In order to treat our problem from the statistical mechanics point of view, for each linked pair $(i,\alpha)$ (i.e. a pair user-tower connected by an edge in the network), and a given $z$-matching, we introduce variables $s_{i\alpha}\in\{{1,0}\}$ such that $s_{i\alpha}=1$ if the edge is included in the $z$-matching, and $s_{i\alpha}=0$ otherwise. 
A $z$-matching reduces to an assignment $\{s_{i\alpha}\}$ that satisfies the conditions
\bea 
\sum_{\alpha\in{N(i)}}s_{i\alpha}&\leq{1},\nonumber \\
\sum_{i\in{N(\alpha)}}s_{i\alpha}&\leq{z},
\eea
where $N(i)$ denotes the set of neighbours of a node $i$ and $N(\alpha)$ denotes the set of neighbours of a tower $\alpha$.
Let us  define the {\it energy} $E$ and the {\it capacity} $C$ of the $z$-matching as 
\bea
E&=&\sum_{i=1}^{N}{E_{i}}+\sum_{\alpha=1}^{M}{E_{\alpha}},\nonumber\\
C&=&\sum_{i=1}^N (1-E_i), \label{eq energy}
\eea
with $E_i$ and $E_{\alpha}$ given by 
\bea 
E_{i}&=&{1-\sum_{\alpha\in{N(i)}}s_{i\alpha}},\nonumber \\
E_{\alpha}&=&z-\sum_{i\in{N(\alpha)}}s_{i\alpha}.
\eea
Therefore, the capacity corresponds to the number of users with one matched link.
Having in mind that $\sum_{i=1}^{N}\sum_{\alpha\in{N(i)}}s_{i\alpha}=\sum_{\alpha=1}^{M}\sum_{i\in N(\alpha)}s_{i\alpha}$, we have the following simple relationship
\bea
E=(zM+N)-2C.
\label{EC}
\eea

The problem of finding a maximum capacity $C$ of the $z$-matching translates then to the problem of investigating allowed configurations $\{{s_{i\alpha}}\}$ for the $z$-matching which minimise the energy $E$. 

Here we associate to each possible $z$-matching of the network the Gibbs measure
\bea \label{eq Gibbs measure}
\hat{P}\left(\{{s_{i\alpha}}\}\right)=\frac{e^{-\beta{E}}}{Z}{\prod_{i=1}^{N}{\theta({1-\sum_{\alpha\in{N(i)}}{s_{i\alpha}}})}}\cdot{\prod_{\alpha=1}^{M}{\theta(z-\sum_{i\in{N(\alpha)}}s_{i\alpha})}},\nonumber
\eea
where $\beta>0$ denotes the inverse temperature, $Z$ is a normalisation constant and $\theta(x)=1$ for $x\geq{0}$ and $\theta(x)=0$ for $x<0$. 
The free-energy of the problem $F(\beta)$ is defined as 
\bea 
\beta F(\beta)=-\ln Z,
\eea
and   the energy $E$ is given by 
\bea
E=\frac{\partial [\beta F(\beta)]}{\partial \beta}.
\label{Energy_f}
\eea
In order to characterise the maximum capacity of a network, or equivalently its minimum energy, we are interested in the limit when $\beta\to {\infty}$.

\section{The Belief Propagation solution on a single network} \label{Sect BP}
\subsection{The Belief Propagation equations}

On a locally tree like bipartite networks, such as a random bipartite network,  the Gibbs measure given by Eq. (\ref{eq Gibbs measure}) can be found in  the Bethe approximation using the Belief Propagation (BP) algorithm \cite{Mezard_Montanari}. The BP algorithm expresses the Gibbs measure in terms of messages. Here we distinguish between two types of messages: those that the users send to neighbouring towers ${P}_{i\to{\alpha}}(s_{i\alpha})$, and those that the towers send to neighbouring users ${P}_{{\alpha}\to i}(s_{i\alpha})$. For a user $i$ and its neighbour $\alpha$, ${P}_{i\to{\alpha}}(1)$ denotes the probability that $i$ says to $\alpha$ that it believes that their edge should be included to the $z$-matching. Similarly, ${P}_{{\alpha}\to{i}}(1)$ denotes the probability that $\alpha$ informs $i$ that their edge should be matched. Then, for $s_{i\alpha}\in\{{0,1}\}$ the Belief Propagation messages are given by
\bea 
{P}_{i\to{\alpha}}(s_{i\alpha})&=&\frac{1}{{\mathcal C}_{i\to \alpha}}\sum_{\bm{s}_{i}\setminus s_{i \alpha}} e^{-\beta E_{i}}  \theta(1-\sum_{\gamma\in N(i)}{s_{i \gamma}})  \nonumber \\
&&\times\prod_{\gamma\in{N(i)\setminus\alpha}} {P}_{\gamma\to{i}}(s_{i \gamma}),\nonumber \\
{P}_{{\alpha}\to{i}}(s_{i\alpha})&=&\frac{1}{{\mathcal C}_{\alpha\to i}}\sum_{\bm{s}_{\alpha}\setminus s_{i \alpha}} e^{-\beta E_{\alpha}}  \theta(z-\sum_{j\in N(\alpha)}{s_{j \alpha}}) \nonumber \\
&&\times\prod_{j \in {N(\alpha)\setminus i}}{P}_{j\to{\alpha}}(s_{j\alpha}),
\label{eq BP}
\eea
where $\bm{s}_i=\{s_{i\alpha}| \alpha\in N(i)\}$, $\bm{s}_\alpha=\{s_{i\alpha}| i\in N(\alpha)\}$, and ${\mathcal C}_{i\to \alpha}$, ${\mathcal C}_{\alpha\to i}$, represent normalisation constants. 
The messages ${{P}_{i\to\alpha}({s}_{i \alpha})},{{P}_{\alpha\to i}({s}_{i\alpha})}$ can be parametrized by the cavity fields $h_{i\to\alpha},\hat{h}_{\alpha\to i}$ in the following way
\bea
{{P}_{i\to\alpha}({s}_{i\alpha})}&=&\frac{e^{\beta  h_{i\to\alpha}  s_{i\alpha}}}{1+e^{\beta  h_{i\to\alpha}}},\nonumber \\
{{P}_{\alpha\to i}({s}_{i\alpha})}&=&\frac{e^{\beta \hat{h}_{\alpha\to i}  s_{i\alpha}}}{1+e^{\beta  \hat{h}_{\alpha \to i}}}.
 \label{eq_cavity_fields}
\eea
By using this parametrization of the messages the BP equations can be written explicitly as
\bea 
{P}_{i\to{\alpha}}(0)&=&\frac{1}{{\mathcal C}_{i\to \alpha}}\left(e^{-\beta} + \sum_{\gamma\in {N(i)\setminus \alpha}}e^{\beta\hat{h}_{\gamma\to i}}\right)\nonumber\\
&&\times\prod_{\gamma\in{N(i)\setminus\alpha}} {P}_{\gamma\to{i}}(0),\nonumber \\
{P}_{i\to{\alpha}}(1)&=&\frac{1}{{\mathcal C}_{i\to \alpha}}\prod_{\gamma\in{N(i)\setminus\alpha}} {P}_{\gamma\to{i}}(0),\nonumber \\
{P}_{{\alpha}\to{i}}(0)&=&\frac{1}{{\mathcal C}_{\alpha\to i}} \nonumber \\
&&\times \left(\sum_{p=0}^{z}e^{-\beta(z-p)}\sum_{j_{1},\dots,j_{p}}e^{\sum_{m=0}^{p}{\beta h_{j_{m}\to\alpha}}}\right) \nonumber\\
&&\times\prod_{j \in {N(\alpha)\setminus i}}{P}_{j\to{\alpha}}(0),\nonumber\\
{P}_{{\alpha}\to{i}}(1)&=&\frac{1}{{\mathcal C}_{\alpha\to i}} \nonumber \\
&& \times \left(\sum_{p=0}^{z-1}e^{-\beta(z-1-p)}\sum_{j_{1},\dots,j_{p}}e^{\sum_{m=0}^{p}{\beta h_{j_{m}\to\alpha}}}\right) \nonumber \\
&&\times\prod_{j \in {N(\alpha)\setminus i}}{P}_{j\to{\alpha}}(0).
\label{eq BP developed}
\eea
Using Eqs. $(\ref{eq BP developed})$ and Eqs. $(\ref{eq_cavity_fields})$, the BP Eqs. (\ref{eq BP}) can find closed expression for the  cavity fields given by  
\bea
&&e^{-\beta{h_{i \to \alpha}}}=e^{-\beta}+\sum_{\gamma \in N(i)\setminus\alpha}{e^{\beta \hat{h}_{\gamma \to i}}},\\
&&e^{-\beta{\hat{h}_{\alpha \to i}}}=e^{-\beta}\nonumber \\
&&+\frac{\sum_{j_{1},\dots,j_l\dots, j_{z}| j_l\in N(\alpha)\setminus i}e^{\sum_{l=0}^{z}\beta{h_{j_{l}\to{\alpha}}}}}{\sum_{p=0}^{z-1}e^{-\beta(z-p-1)}\cdot\sum_{j_{1},\dots,j_l\dots,j_{p}| j_l\in N(\alpha)\setminus i}e^{\sum_{l=0}^{p}{h_{j_{l}\to\alpha}}}}.\nonumber
\eea

\subsection{The marginal probabilities}

According to the BP algorithm \cite{Mezard_Montanari}, the marginal probability ${P}_{i \alpha}(s_{i \alpha})$ of the variable $s_{i \alpha}$ associated with the link between $i$ and $\alpha$ is given by 
\bea
{P}_{i \alpha}(s_{i \alpha})&=&\frac{1}{{\mathcal C}_{i \alpha}}{P}_{i\to{\alpha}}(s_{i\alpha}){P}_{{\alpha}\to{i}}(s_{i\alpha}),
\eea
where ${\mathcal C}_{i \alpha}$ are normalisation constants.

Similarly the marginal probability ${P}_{i}(\bm{s}_{i})$ of the variables  $\bm{s}_i=\{s_{i\alpha}| \alpha\in N(i)\}$ associated to the links incident to node $i$ and the marginal probability 
 ${P}_{\alpha}(\bm{s}_{\alpha})$ of the variables $\bm{s}_\alpha=\{s_{i\alpha}| i\in N(\alpha)\}$  associated to links incident to tower $\alpha$ are given by 
\bea
{P}_{i}(\bm{s}_{i})=\frac{e^{-\beta E_{i}}}{{\mathcal C}_{i}} \theta(1-\sum_{\alpha\in{N(i)}}s_{i\alpha}) \prod_{\alpha \in N(i)}{P}_{\alpha \to i}(s_{i \alpha}),\nonumber \\
{P}_{\alpha}(\bm{s}_{\alpha})=\frac{e^{-\beta E_{\alpha}}}{{\mathcal C}_{\alpha}}  \theta(z-\sum_{i\in{N(\alpha)}}s_{i\alpha}) \prod_{i \in N(\alpha)} {P}_{i \to \alpha}(s_{i \alpha}).
\eea
In the Bethe approximation, valid on locally tree-like networks the Gibbs measure given by Eq. (\ref{eq Gibbs measure}) is given in terms of the marginals by
\bea
\hat{P}\left(\{s_{i \alpha}\}\right)=\frac{\prod_{i=1}^{N}{ P}_{i}(\bm{s}_{i})\prod_{\alpha=1}^{M}{P}_{\alpha}(\bm{s}_{\alpha})}{\prod_{(i,\alpha)} {P}_{i\alpha}({s}_{i \alpha})}.
\label{P_Bethe}
\eea

\subsection{Free energy}

The free energy of the system can be found by minimising the Gibbs free energy given by 
\bea
\beta F_{Gibbs}=\sum_{\{s_{i\alpha}\}}\hat{P}\left(\{s_{i \alpha}\}\right)\ln\left(\frac{\hat{P}\left(\{s_{i \alpha}\}\right)}{\psi(\{s_{i\alpha}\})}\right),
\label{Gibbs free energy}
\eea
where 
\bea
\psi\left(\{{s_{i\alpha}}\}\right)={e^{-\beta{E}}}{\prod_{i=1}^{N}{\theta({1-\sum_{\alpha\in{N(i)}}{s_{i\alpha}}})}}\cdot{\prod_{\alpha=1}^{M}{\theta(z-\sum_{i\in{N(\alpha)}}s_{i\alpha})}}.
\eea
In fact it can easily be shown that the Gibbs free energy is minimised when $
\hat{P}\left(\{{s_{i\alpha}}\}\right)$ is given by Eq. (\ref{eq Gibbs measure}) and that the minimum Gibbs free-energy is equal to the free energy of the problem and takes the value
\bea
\beta F_{Gibbs}=\beta F(\beta)=-\ln Z.
\eea
Using the Bethe expression for the Gibbs measure given by Eq. (\ref{P_Bethe}) in  Eq. (\ref{Gibbs free energy}) we obtain the free energy
\bea
&&\beta F({\beta})=\sum_{(i,\alpha)}\ln{\mathcal C}_{i \alpha}-\sum_{i=1}^{N}\ln{\mathcal C}_{i}-\sum_{\alpha=1}^{M}\ln{\mathcal C}_{\alpha} \\
&&=\sum_{(i,\alpha)}\ln\left({1+e^{\beta(h_{i\to\alpha}+\hat{h}_{\alpha\to i})}}\right)\nonumber \\
&&-\sum_{i=1}^N\ln{\left(e^{-\beta}+\sum_{\gamma\in{N(i)}}e^{\beta {\hat{h}_{\gamma\to i}}}\right)}\nonumber \\
&&-\sum_{\alpha=1}^{M}\ln{\left({\sum_{p=0}^{z}{e^{-\beta(z-p)}}}\sum_{j_1\dots,j_l\dots,j_{p}| j_l\in N(\alpha)}e^{\beta{\sum_{l=1}^{p}h_{j_{l}\to{\alpha}}}}\right)}.\nonumber
\eea

Finally, using Eq. (\ref{Energy_f}) we can express the energy $E$ in terms of the cavity fields as 
\bea
&&E=\sum_{(i,\alpha)}\frac{e^{\beta (h_{i\to\alpha}+\hat{h}_{\alpha\to i})}(h_{i\to\alpha}+\hat{h}_{\alpha\to i})}{1+e^{\beta (h_{i \to \alpha}+\hat{h}_{\alpha \to i})}}\nonumber \\
&&+\sum_{i=1}^{N}\frac{e^{-\beta}-\sum_{\gamma\in{N(i)}}e^{\beta \hat{h}_{\gamma\to i}} \hat{h}_{\gamma\to i}}{e^{-\beta}+\sum_{\gamma\in N(i)}e^{\beta \hat{h}_{\gamma\to i}}}\nonumber \\
&&\hspace{-15mm}+\sum_{\alpha=1}^{M}\frac{\sum_{p=0}^{z}\sum_{j_{1},\dots,j_{p}}e^{-\beta(z-p-\sum_{l=1}^{p}h_{j_{l}\to \alpha})}g(\{h_{\ell\to\alpha}\}) }{\sum_{p=0}^{z}e^{-\beta(z-p)}\sum_{j_{1},\dots,j_{p}}e^{\beta\sum_{l=1}^{p}{h_{j_{l}\to\alpha}}}},
\eea
where 
\bea
g(\{h_{\ell\to\alpha}\})=\left[(z-p)- \theta(p)\sum_{l=1}^{p}h_{j_{l}\to\alpha}\right].
\eea
\subsection{The zero temperature limit}

For finding the maximum capacity of a bipartite network we need to investigate the zero temperature limit of the BP equations, i.e. we need to consider the limit $\beta\to \infty$. In this limit the cavity fields $h_{i\alpha}, \hat{h}_{\alpha\to i}$ have the support on $\{-1,0,1\}$ and the BP equations are 
\bea 
h_{i\to\alpha}&=&-\max\left[-1,\max_{\gamma\in N(i)\setminus\alpha} \hat{h}_{\gamma\to i} \right],\nonumber \\
\hat{h}_{\alpha\to i}&=&\left\{\begin{array}{ll}
1 & \mbox{if } \sum_{j\in N(\alpha)\setminus i}\delta(-1,h_{j\to \alpha})\geq q-z,\\
-1 & \mbox{if } \sum_{j\in N(\alpha)\setminus i}\delta(1,h_{j\to \alpha})\geq z,\\
0  & \mbox{otherwise},
\end{array}\right.
\label{eq BP zero temp}
\eea
where $\delta(x,y)$ denotes the Kronecker delta.
Thus, a node $i$ sets a field $h_{i \to \alpha}=1$ if all other neighbouring towers set the fields which point to $i$ to $-1$; it sets $h_{i \to \alpha}=-1$ if at least one other neighbouring tower sends $+1$; and it sets $h_{i \to \alpha}$ to $0$ otherwise, i.e. if at least one other tower sends $0$, and no tower sends $+1$. Similarly, a tower $\alpha$ of degree $q\geq{z}$ sets a field $h_{\alpha \to i}$ to $1$ if at least $q-z$ of its neighbouring users set fields pointing to $\alpha$ to $-1$; $\alpha$ sets the field  $h_{\alpha \to i}$ to $-1$ if at least $z$ neighbours set fields pointing to it to $+1$; and otherwise, it sets the field $h_{\alpha \to i}$ to $0$.

In the case of multiple solutions, the dynamically stable solution that is physical and that minimises the energy $E$ which can be expressed as
\bea \label{eq energy zero temp}
E&=&\sum_{(i,\alpha)}\max\left[0,h_{i\to\alpha}+\hat{h}_{\alpha\to i}\right]\nonumber \\
&&-\sum_{i=1}^{N}\max\left[-1,\max_{\gamma\in N(i)}\hat{h}_{\gamma\to i}\right]\nonumber \\
&&-\sum_{\alpha=1}^{M}\max\left[-z,\max_{{\{j_{l}|j_l\in N(\alpha)\}_{l\leq z}}}\sum_{l=1}^{z}h_{j_{l}\to\alpha}\right],
\eea
represents the solution of the maximum capacity problem.
We note that the equations obtained for the $z$-matching problem by considering the limit $\beta\to \infty$ of the BP equations clearly reduce to the equations obtained in \cite{Lenka_Mezard} for the matching problem when $z=1$.

\section{Maximum $z$-matching problem on bipartite network ensembles} \label{Sect Ensambles}
\subsection{The BP equations and the energy of the $z$-matching}
Here we consider the maximum $z$-matching problem on bipartite network ensembles formed by $N$ users and $M$ towers where the users have degree distribution $\tilde{P}^{(U)}(k)$, and the towers have degree distribution $\tilde{P}^{(T)}(q)$, with $\tilde{P}^{(T)}(q)=0$ for $q<z$. Note that on a bipartite network  the average degree $\avg{k}$ of the users and the average degree $\avg{q}$ of the towers need to satisfy
\bea
N\avg{k}=M\avg{q}.
\label{eq bipartite avg degree}
\eea
In order to study the energy $E$ of the $z$-matching problem on these ensembles we denote by 
 ${\mathcal{P}}^{(U)}(h)$ and ${\mathcal{P}}^{(T)}(\hat{h})$ the distributions of fields $h$ and $\hat{h}$, respectively, i.e.
\bea
{\mathcal{P}}^{(U)}(h)&=&w_{1} \delta(h,1)+w_{2} \delta(h,-1)+w_{3} \delta(h,0),\nonumber \\
{\mathcal{P}}^{(T)}(\hat{h})&=&\hat{w}_{1} \delta(\hat{h},1)+\hat{w}_{2} \delta(\hat{h},-1)+\hat{w}_3 \delta(\hat{h},0),
\eea
where $w_{1}+w_{2}+w_{3}=1$ and $\hat{w}_{1}+\hat{w}_{2}+\hat{w}_{3}=1$.
Therefore, $w_1, w_2,w_3 $ indicate the probability that the cavity fields $h$ coming from users are  equal to $1,-1,0$ respectively, and $\hat{w}_1,\hat{w}_2,\hat{w}_3$ indicate the probability that the cavity fields $\hat{h}$ coming from towers are equal to $1,-1,0$, respectively.

Using the BP Eqs. (\ref{eq BP zero temp}) derived in the zero temperature limit, we can derive the equations satisfied by the probabilities $\{w_1,w_2,w_3\}$ in a bipartite network ensemble. We have 
\bea 
w_{1}&=&\sum_{k}{\frac{k \tilde{P}^{(U)}(k)}{\avg{k}} \hat{w}_2^{k-1}},\nonumber\\
w_{2}&=&1-\sum_{k}{\frac{k \tilde{P}^{(U)}(k)}{\avg{k}} (1-\hat{w}_{1})^{k-1}},\nonumber \\
w_3&=&1-w_1-w_2.
\label{eq cavity field weights users}
\eea
This is consistent with the previous discussion. Namely, a node of degree $k$ sets the outgoing field to $1$ if all remaining $k-1$ neighbours set corresponding incoming fields to $-1$; it sets the outgoing field to $-1$ if at least one of the remaining $k-1$ neighbours sets the incoming filed to $1$; otherwise, it sets it to $0$.
Similarly, we can derive the equations satisfied by probabilities $\{\hat{w}_1,\hat{w}_2,\hat{w}_3\}$ in the bipartite network ensemble,
\bea 
\hat{w}_1&=&\sum_{q}\frac{q \tilde{P}^{(T)}(q)}{\avg{q}} \sum_{p=q-z}^{q-1}{q-1 \choose p}w_{2}^p  (1-w_{2})^{q-1-p},\nonumber\\
\hat{w}_{2}&=&1-\sum_{q}\frac{q \tilde{P}^{(T)}(q)}{\avg{q}}\sum_{p=0}^{z-1} {q-1 \choose p} w_{1}^p  (1-w_{1})^{q-1-p},\nonumber \\
\hat{w}_3&=&1-\hat{w}_1-\hat{w}_2.
\label{eq cavity field weights towers}
\eea
Again, as previously discussed, a tower of degree $q\geq{z}$ sets the outcoming field to $+1$ if at least $q-z$ of remaining $q-1$ neighbours set corresponding incoming fields to $-1$; it sets the field to $-1$ if at least $z$ neighbours set the corresponding fields to $+1$; otherwise, it sets the field to $0$.
Finally, the energy $E$ of the maximum $z$-matching Eq. $(\ref{eq energy zero temp})$ can be expressed in terms of the probabilities $\{w_{1},w_{2},w_{3},\hat{w}_{1},\hat{w}_{2}, \hat{w}_{3}\}$ as 
\bea
E&=&N\sum_{k}{\tilde{P}^{(U)}(k)}\left[\hat{w}_{2}^k-\left(1-\left(1-{\hat{w}_{1}}\right)^k\right)\right]\nonumber \\&&\hspace*{-8mm}-M\sum_{q}{\tilde{P}^{(T)}(q)} \Bigg[z\sum_{p=z}^{q} {q \choose p}{w_{1}}^p\left(1-{w_{1}}\right)^{q-p} \nonumber\\
&&\hspace*{-8mm}+\sum_{p_{1}=0}^{z-1}\sum_{p_{3}=z-p_{1}}^{q-p_{1}}\frac{q!}{p_1!p_3!(q-p_1-p_3)!}p_{1}w_{1}^{p_{1}}{w_{3}}^{p_{3}}{w_{2}}^{q-p_{1}-p_{3}}\nonumber \\&&\hspace*{-8mm}+\sum_{p_{1}=0}^{z-1}\sum_{p_{3}=0}^{z-p_{1}-1}\frac{q! \ (2p_{1}+p_{3}-z)}{p_1!p_3!(q-p_1-p_3)!}w_{1}^{p_{1}}w_{3}^{p_{3}}w_{2}^{q-p_{1}-p_{3}}\Bigg]\nonumber \\
&&\hspace*{-8mm}+N\avg{k}\left[w_{1}\left(1-\hat{w}_{2}\right)+\hat{w}_{1}\left(1-w_{2}\right)\right].
\label{eq energy ensemble}
\eea
%where 
%\bea
%\hat{g}(\{p_m\})=2p_{1}+p_{3}-z.
%\eea

Therefore, the phase diagram of the $z$-matching problem can be drawn by solving Eqs. $(\ref{eq cavity field weights users})$ and Eqs. $(\ref{eq cavity field weights towers})$ and calculating the energy $E$ given by Eq. (\ref{eq energy ensemble}) on this solution as a function of the structural properties of the bipartite network ensemble.
Finally we  note that the equations obtained here for  the $z$-matching problem defined on bipartite network ensembles reduce to the equations obtained in \cite{Lenka_Mezard} for the matching problem in the limit case $z=1$.

\subsection{Stability condition }

The solutions of the BP Eqs. $(\ref{eq cavity field weights users})$ and $(\ref{eq cavity field weights towers})$ should be physical, i.e. they should correspond to values of the capacity
\bea
0\leq C\leq \min(zM,N).
\label{physical}
\eea
Moreover, they should be dynamically stable. In order to characterise the stability of a given solution we calculate the  Jacobian matrix $J$ of the system of equations for the probabilities $\{w_{1},w_{2},w_{3},\hat{w}_{1},\hat{w}_{2}, \hat{w}_{3}\}$ including Eqs. $(\ref{eq cavity field weights users})$ and Eqs. $(\ref{eq cavity field weights towers})$  which is
\bea
J=\left(\begin{array}{cccccc}
0 & 0 & 0 & 0 & G'_{1,k}(\hat{w}_2) & 0\\
0 & 0 & 0 & G'_{1,k}(1-\hat{w}_{1}) & 0 & 0\\
-1 & -1 & 0 & 0 & 0 & 0\\
0 & A(w_2) & 0 & 0 & 0 & 0 \\
B(w_1) & 0 & 0 & 0 & 0 & 0\\
0 & 0 & 0 & -1 & -1 & 0
\end{array}\right),\nonumber
\eea
where
\bea
G'_{1,k}(x)=\sum_{k}\frac{k(k-1)}{\avg{k}}\tilde{P}^{(U)}(k)x^{k-2},
\eea
and %where $A(w_2),B(w_1)$ are given by  
\bea
A(w_2)&=&\sum_{q} \frac{q \tilde{P}^{(T)}(q)}{\avg{q}}\sum_{p=0}^{z-1} {q-1 \choose p} H_A(\{w_m\}), \nonumber \\
B(w_1)&=&\sum_{q} \frac{q \tilde{P}^{(T)}(q)}{\avg{q}}\sum_{p=0}^{z-1} {q-1 \choose p} H_B(\{w_m\}),\nonumber 
\eea
with 
\bea
\hspace*{-5mm}&&H_A(\{w_m\})=\left[(q-1-p)-(q-1)w_{2}\right] w_{2}^{q-2-p}(1-w_{2})^{p-1},\nonumber \\
\hspace*{-5mm}&&H_B(\{w_m\})=\left[(q-1)w_{1}-p\right]w_{1}^{p-1}(1-w_{1})^{q-2-p},\nonumber
\eea
and in the derivation of $A(w_2)$ we use 
\bea\sum_{p=q-z}^{q-1}{q-1 \choose p}w_{2}^{p}(1-w_2)^{q-1-p}=\sum_{p=0}^{z-1}{q-1 \choose p}w_{2}^{q-1-p}(1-w_2)^{p}.\nonumber\eea
A given solution of the system of Eqs. (\ref{eq cavity field weights users}) and Eqs. (\ref{eq cavity field weights towers}) is stable if and only if eigenvalues of the Jacobian $J$ are all less than one. In this way we obtain the stability conditions
\bea
B(w_1) G'_{1,k}(\hat{w}_{2})<1,\nonumber \\
A(w_2) G'_{1,k}(1-\hat{w}_{1})<1.\label{eq stab cond}
\eea

The trivial solution $w_{1}=w_{2}=\hat{w}_{1}=\hat{w}_{2}=0$ and $w_3=\hat{w}_3=1$ deserves a special consideration. This solution corresponds to $E=0$, i.e. capacity $C=(N+zM)/2$ (recall Eq. (\ref{EC})). It immediately follows that this solution is physical, i.e. it satisfies Eq. $(\ref{physical})$ only for 
\bea
N=zM,
\label{condition trivial}
\eea in which case it corresponds to full capacity
\bea
C=N=zM.
\eea
From the study of the  BP Eqs. $(\ref{eq cavity field weights users})$ and $(\ref{eq cavity field weights towers})$  we observe that  these equations admit the trivial solution $w_{1}=w_{2}=\hat{w}_{1}=\hat{w}_{2}=0$ and $w_3=\hat{w}_3=1$ as long as the minimum degree of the nodes is two and the minimum degree of the towers is $z+1$, i.e.  $\tilde{P}^{(U)}(k)=0$ for $k=0,1$ and $\tilde{P}^{(T)}(q)=0$ for $q\leq z$. However, in order to establish whether this is the solution of the maximum $z$-matching problem, we need to investigate its stability. 

In particular, if we study the stability conditions Eqs. (\ref{eq stab cond}) of the trivial solution $w_{1}=w_{2}=\hat{w}_{1}=\hat{w}_{2}=0$ and $w_3=\hat{w}_3=1$, we obtain
\bea
\frac{\Avg{q(q-1)}}{\avg{q}}\frac{2\tilde{P}^{(U)}(2)}{\avg{k}}&<1,\nonumber \\
\frac{(z+1){z}\tilde{P}^{(T)}(z+1)}{\avg{q}}\frac{\Avg{k(k-1)}}{\avg{k}}&<1.
\eea
Therefore, the instability of the trivial solution on a bipartite network ensemble with given degree distribution is driven by the fraction of users of degree two and the fraction of towers of degree $z+1$.
In particular when the minimum degree of the nodes is greater than two, i.e. $\tilde{P}^{(U)}(1)=\tilde{P}^{(U)}(2)=0$ and the minimum degree of the towers is greater than $z+1$, i.e. $\tilde{P}^{(T)}(q)=0$ for $q\leq z+1$, as long as $N=zM$ we get that the trivial solution is stable independently of the other properties of the degree distributions of the nodes and of the towers. This  generalises the relation found in matching of simple networks\cite{Lenka_Mezard}, in matching of directed networks \cite{Menichetti_PRL} and on generalised matching in multilayer networks \cite{Menichetti_multilayer}.

\subsection{$z$-matching in network ensembles}

Let us discuss here few examples of the phase diagram of the maximum $z$-matching on bipartite networks ensembles.
Let us start with a simple example of a regular bipartite network in which the degree distributions are given by 
\bea
\tilde{P}^{(U)}(k)&=&\delta(k,\bar{k}),\nonumber \\
\tilde{P}^{(T)}(q)&=&\delta(q,\bar{q}),
\eea
with $\bar{k}>0$ and $\bar{q}\geq z$ where $\delta(x,y)$ denotes the Kronecker delta.
For these networks we clearly have $\bar{k}=\avg{k}$ and $\bar{q}=\avg{q}$, therefore it follows that Eq. (\ref{eq bipartite avg degree}) reduces to 
\bea
\bar{k}N=\bar{q}M.
\label{kq}
\eea
In order to characterise the $z$-matching on this network ensemble, we solve the Eqs. (\ref{eq cavity field weights users}) together with  Eqs. ($\ref{eq cavity field weights towers}$) for the probabilities $\{w_1,w_2,w_3,\hat{w}_1,\hat{w}_2,\hat{w}_3\}$ and we evaluate the energy $E$ using Eq. (\ref{eq energy ensemble}) and hence the capacity $C$ using Eq. (\ref{EC}).

As  a function of the values of $\bar{k}$ and $\bar{q}$ we have the following regimes:
\begin{itemize}
\item[(1)] If $z\bar{k}>\bar{q}$, or equivalently $zM>N$  the solution is $w_2=\hat{w}_1=1$ and $w_{1}=w_3=\hat{w}_{2}=\hat{w}_{3}=0$ and  corresponding to energy $E=-N+zM$ and capacity $C=N$.
\item[(2)] If $z\bar{k}=\bar{q}$ or equivalently $zM=N$ the solution is $w_{3}=\hat{w}_{3}=1$ and $w_1=w_2=\hat{w}_1=\hat{w}_2=0$ for $\bar{k}>1$ and $w_{1}=\hat{w}_{1}=1$, $w_2=w_3=\hat{w}_2=\hat{w}_3=0$ for $\bar{k}=1$. Both solutions correspond to energy $E=0$ and capacity $C=N$.
\item[(3)] If $z\bar{k}<\bar{q}$ or equivalently $zM<N$ the solution is $w_{1}=\hat{w}_{2}=1$ and $w_2=w_3=\hat{w}_1=\hat{w}_3=0$ corresponding to energy $E=N-zM$ and capacity $C=zM$.
\end{itemize}

\begin{figure}
  \includegraphics[width=\columnwidth]{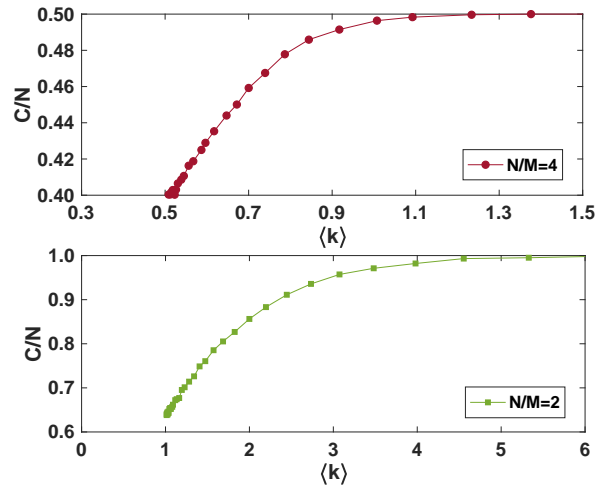}
\caption{
Capacity density $C/N$ of the $z$-matching of a bipartite network with Poisson degree distribution of the nodes and of the towers given by Eqs. (\ref{poi}) is plotted versus $\Avg{k}$ for $z=2$ at constant value of $N/M$. The results are obtained using the message passing algorithm averaged over 70 bipartite networks realisations.}
\label{fig:poisson}      
\end{figure}

As a second example of bipartite network ensemble we consider the bipartite network in which the degree distributions of the towers and the nodes are Poisson distributed according to the distribution
\bea
\begin{array}{ccc}
\tilde{P}^{(U)}(k)\sim\frac{a^k}{k!}e^{-a} &\mbox{with} &0\leq k\leq M,\\
\tilde{P}^{(T)}(q)\sim\frac{b^{q-z}}{(q-z)!}e^{-b} &\mbox{with} &z\leq q\leq N,
\end{array}
\label{poi}
\eea 
where $a,b$ are related so that Eq. $(\ref{eq bipartite avg degree})$ holds.
 For infinite networks with   the degree distribution of  users   $\tilde{P}^{(U)}(k)$ and of towers $\tilde{P}^{(T)}(q)$ given by  Eq. (\ref{poi}) the full capacity solution is never achieved as we will always have $\tilde{P}^{(U)}(1)>0$ and $\tilde{P}^{(T)}(z)>0$. 
 However for finite networks, the fraction of users with degree 1 and the fraction of towers with degree $z$ is likely to be zero if $\tilde{P}^{(U)}(1)<1/N$ and  $\tilde{P}^{(T)}(z)<1/M$. Thus, in this case it is possible to enter the regime of the trivial solution which guarantees the full capacity. 
 In Figure \ref{fig:poisson} we show plot the average capacity density $C/N$ versus the average degree  $\Avg{k}$ for bipartite networks having different ratio $N/M$ between the number of towers and the number of users. We see that for sufficiently high average degree $\avg{k}$ the full capacity solution can be achieved. In particular we observe that as the average degree increases the $z$-mathcing problem converge to the full capacity solution $C/N=zM/N$ for $zM\leq N$. The results are obtained averaging the results obtained using the message passing algorithm on single realizations of the bipartite networks with degree distributions  of  users   $\tilde{P}^{(U)}(k)$ and of towers $\tilde{P}^{(T)}(q)$ given by  Eq. (\ref{poi}) over 70 network realizations.

\section{Conclusions} \label{Sect Conclusion}

In this paper we have analysed the many-to-one $z$-matching problem using a statistical mechanics approach. This problem is inspired by a wireless network scenario where a set of users needs to be matched  to a set of towers providing the wireless connection. While a user can be connected at most with one tower, a tower can serve up to $z$ users. Here we have used the Belief Propagation algorithm in the zero temperature limit to characterise the bipartite network capacity, i.e. the fraction of matched users which is a good proxy for the efficiency of the communication in the network. The phase-diagram of the $z$-matching problem has been derived for different bipartite network ensembles with given degree distribution. 

As the matching problem has recently been related to the controllability of the network, in the future we plan to explore whether also the $z$-matching problem can be related to the dynamics defined on bipartite networks.

\acknowledgments
We acknowledge interesting discussions with Alex Kartun-Giles.

\end{document}